\documentclass[12pt,a4paper,fleqn]{article}
\usepackage[utf8]{inputenc}
\usepackage{authblk}
\usepackage{geometry}
\usepackage{xcolor}
\usepackage{natbib}
\usepackage{graphicx}
\usepackage{pdflscape}
\usepackage{appendix}
\usepackage{setspace}
\usepackage{makecell}
\usepackage{graphbox}

\usepackage{colortbl}

\usepackage{threeparttable}
\usepackage{booktabs}

\usepackage{amssymb}
\usepackage{amsbsy}
\usepackage{bm}
\usepackage{amsmath}
\usepackage{amsthm}
\usepackage{graphicx}
\usepackage{mathtools}

\usepackage{enumitem}
\newlist{steps}{enumerate}{1}
\setlist[steps, 1]{label = Step \arabic*:}

\usepackage{dcolumn}
\newcolumntype{d}[1]{D{.}{.}{#1}}



\usepackage{caption}
\usepackage{subcaption}
\definecolor{nblue}{HTML}{000660}
\usepackage[colorlinks=true,urlcolor=nblue,linkcolor=nblue,citecolor=nblue]{hyperref}
\addtocounter{page}{-1}

\title{\LARGE \textbf{Machine Learning the Macroeconomic Effects of Financial Shocks}\thanks{
\noindent Corresponding author: Karin Klieber. Monetary Policy Section, Oesterreichische Nationalbank. \textit{Address}: Otto-Wagner-Platz 3, 1090 Vienna, Austria. \textit{Email}: \href{mailto:karin.klieber@oenb.at}{karin.klieber@oenb.at}. The authors thank Serena Ng for helpful comments and suggestions.  Hauzenberger and Huber gratefully acknowledge financial support from the Jubiläumsfonds of the Oesterreichische Nationalbank (OeNB, grant no. 18763) and the Austrian Science Fund (FWF, grant no. ZK 35). Marcellino thanks for funding the European Union - NextGenerationEU, Mission 4, Component 2, in the framework of the GRINS -Growing Resilient, INclusive and Sustainable project (GRINS PE00000018 – CUP B43C22000760006). The views expressed in this paper do not necessarily reflect those of the Oesterreichische Nationalbank or the Eurosystem or the European Union, and they cannot be held responsible for them.}}
\author[1]{\MakeUppercase{Niko Hauzenberger}}
\author[2]{\MakeUppercase{Florian Huber}}
\author[3]{\MakeUppercase{Karin Klieber}}
\author[4]{\MakeUppercase{Massimiliano Marcellino}}
\affil[1]{\textit{University of Strathclyde}}
\affil[2]{\textit{University of Salzburg}}
\affil[3]{\textit{Oesterreichische Nationalbank}}
\affil[4]{\textit{Bocconi University, IGIER, CEPR, Baffi-Carefin and BIDSA}}

\begin{document}

\maketitle\thispagestyle{empty}\normalsize\vspace*{-1em}\small

\begin{center}
\begin{minipage}{0.8\textwidth}
\noindent\small  We propose a method to learn the nonlinear impulse responses to structural shocks using neural networks, and apply it to uncover the effects of US financial shocks. The results reveal  substantial asymmetries with respect to the sign of the shock. Adverse financial shocks have  powerful effects on the US economy, while benign shocks trigger much smaller reactions. Instead, with respect to the size of the shocks, we find no discernible asymmetries.\\ \\ 

\end{minipage}
\end{center}

\spacing{1.5}\normalsize\renewcommand{\thepage}{\arabic{page}}

\newpage

\section{Introduction}



Extreme financial shocks such as the bankruptcy of Lehman Brothers in September 2008 have the potential to trigger substantial nonlinear reactions on the real side of the economy. These nonlinearities appear in terms of size, sign and state dependence \citep{brunnermeier2014macroeconomic}. These asymmetric effects, however, often arise from assuming particular functional relations for the conditional mean part of the model and are thus inherently model-dependent. In this note, we solve this issue by ``machine learning" the nonlinear effects of financial shocks on US macroeconomic aggregates using Bayesian neural networks. 

Existing studies learn the domestic effects of financial shocks either through flexible nonlinear parametric models \citep{barnichon2022effects} or through nonparametric techniques \citep{mumtaz2022impulse}. In this note,  we propose using Bayesian Neural Networks (BNNs), a model that has, so far, demonstrated its effectiveness in forecasting \citep{hauzenberger2024BNN}. By basing local projections on BNNs, we gain additional flexibility in modeling nonlinearities and address the common criticism of black-box models as being unsuitable for structural economic analysis, demonstrating that BNNs can offer meaningful insights into economic dynamics. 


We use BNNs as developed in \cite{hauzenberger2024BNN} to investigate how key US macroeconomic quantities react to a financial shock, and assess whether the reactions are non-proportional with respect to the size and asymmetric with respect to the sign of the shock. Specifically, we develop BNN-based nonlinear local projections \citep[NLPs, see][]{jorda2005estimation} and investigate how financial shocks --- measured with the excessive bond premium \citep[EBP,][]{gilchrist2012credit} --- impact US inflation, industrial production, and employment.  We find substantial asymmetries with respect to sign, with negative shocks exerting stronger effects than positive ones, but substantial proportionality with respect to the size of the shock.

The note is structured as follows. Section \ref{seclp:bnns} gives a short introduction to Bayesian Neural Networks. Section \ref{seclp:structlp} develops nonlinear local projections for BNNs. Section \ref{seclp:results} presents the propagation of financial shocks via our nonlinear framework. Section \ref{seclp:conclusion} concludes.

\section{Bayesian Neural Networks (BNNs)}\label{seclp:bnns}
We follow \cite{hauzenberger2024BNN} and use Bayesian Neural Networks (BNNs) to introduce nonlinearities into a general regression model given by:
\begin{align}
y_t &= \bm x_t'\bm \gamma + f(\bm x_t) + \varepsilon_t, \quad \varepsilon_t \sim \mathcal{N}(0, \sigma_t^2),\label{eq:reg_general} \\
f(\bm x_t) &\approx \widehat{f}_L(\bm x_t) = \bm W_{L+1} \bm h_L\left( \bm W_L \bm h_{L-1} (\cdots \bm W_2 \bm h_1 (\bm W_1 \bm x_t))\right),\label{eq:DNN}
\end{align}
where $\bm \gamma$ denotes linear coefficients of dimension $K$, and $\varepsilon_t$ is a Gaussian shock with zero mean and time-varying variance $\sigma_t^2$.\footnote{The error variance is modeled via stochastic volatility, as in \cite{kastner2014ancillarity}.} 
We allow for $L$ hidden layers and $Q_\ell~(\ell=1, \dots, L)$ neurons in each layer. By recursively applying nonlinear transformations to the neurons of the previous layer, we move from covariates $\bm x_t$ to $f(\bm x_t)$, the output of the final layer. These transformations are implemented via activation function $h_{\ell, q}~(q=1, \dots, Q_\ell)$, which in the case of \cite{hauzenberger2024BNN} can be layer and neuron-specific (i.e., $\bm h_\ell = (h_{\ell, 1}, \dots, h_{\ell, Q_\ell})'$).\footnote{Note that this discussion abstracts from the bias term to simplify notation. In our empirical application, we include the bias term, which in a neural network, allows the activation function to be shifted towards positive and negative values.} 
The network coefficients are stored in $\bm W_\ell$ for $\ell=2,\dots,L$ with dimension $Q_\ell \times Q_{\ell-1}$, in $\bm W_{L+1}$ as a $1 \times Q_L$ vector and $\bm W_1$ as a $Q_1 \times K$ matrix. 

To address the sharp increase in the number of coefficients with more complex network structures, we introduce regularization via a global local shrinkage prior in the form of the horseshoe prior \citep{carvalho2009handling,ghosh2019model,bhadra2020horseshoe}. In this setup, each element of $\bm w_{\ell, i \bullet} = (w_{\ell, i1}, \dots, w_{\ell, i Q_{\ell -1}})'$, with $\bm w_{\ell, i \bullet}$ denoting the $i^{th}$ row of $\bm W_\ell~(\ell=1, \dots, L+1)$ follows:
\begin{equation*}
    w_{\ell, ij} \sim \mathcal{N}(0, \phi_{{\ell, ij}}), \quad \phi_{{\ell, ij}} = \lambda^2_{{\ell, i}} \varphi^2_{\ell, ij}, \quad \lambda_{{\ell, i}} \sim \mathcal{C}^+(0,1), \quad \varphi_{{\ell, ij}} \sim \mathcal{C}^+(0,1).
\end{equation*}
The global (neuron-specific) shrinkage parameter $\lambda^2_{{\ell, i}}$ forces all elements in $\bm w_{\ell, i \bullet}$ towards zero, while the local scaling parameter $\varphi_{\ell, ij}$ allows for coefficient-specific shrinkage. Accordingly, we apply the horseshoe prior to the linear coefficients $\bm \gamma$.

One important specificity of the network structure is that it allows for a mixture specification, averaging over four different activation functions. We define $h^{(m)}$ as one out of $m \in \{\text{leakyReLU}, \text{sigmoid}, \text{ReLU}, \text{tanh}\}$ activation functions and let each be given by:
\begin{equation}
        h_{\ell, q}(z_{\ell q, t}) = \sum_{m=1}^4  {\omega}^{(m)}_{\ell, q} h^{(m)}(z_{\ell q, t}),
\end{equation}
with $z_{\ell q, t}$ denoting the $q^{th}$ element in the recursively defined vector $\bm z_{\ell, t} = \bm W_{\ell} \bm h_{\ell-1}( \bm z_{\ell-1, t})$ and $\bm z_{1, t} = \bm W_1 \bm x_t$. Weights ${\omega}^{(m)}_{\ell, q}$ are constrained to satisfy ${\omega}^{(m)}_{\ell, q} \ge 0$ and $\sum_m {\omega}^{(m)}_{\ell, q} = 1$. The prior on $\omega^{(m)}_{\ell, q}$ is set in an uninformative manner with a prior probability of $\text{Prob}(\delta_q = m) = 1/4$. 

For posterior inference, we use an MCMC algorithm structured into multiple blocks, iterated 20,000 times with the initial 10,000 draws discarded as burn-in. In brief, we start by drawing the linear coefficients $\bm \gamma$ and the nonlinear coefficients of the last layer $\bm W_{L+1}$ jointly from a standard multivariate Gaussian posterior. The remaining coefficients $\bm W_\ell| \bullet$, for $\ell = 1, \dots, L$ are obtained via an HMC step \citep{neal2011hmc}. All shrinkage hyperparameters are updated by sampling from inverse Gamma distributions using the sampler as in \cite{makalic2015simple}. Finally, to simulate the activation function $h_{\ell, q}$, we draw the indicator $\delta_{\ell, q}$ from a multinomial distribution. $\delta_{\ell, q}$ takes integer values from one to four, each corresponding to a specific activation function selected from the predefined set. For technical details we refer to \cite{hauzenberger2024BNN}.

\section{Nonlinear local projections in BNNs}\label{seclp:structlp}
In the literature, there is substantial evidence of asymmetric effects of benign versus adverse financial shocks on the economy \citep[see, e.g.,][]{balke2000credit,brunnermeier2014macroeconomic,barnichon2022effects}. To shed light on this issue, we develop nonlinear local projections \citep[NLPs; see, e.g.,][]{mumtaz2022impulse, gonccalves2024state, inoue2024local} for  BNNs. 

Let $\zeta_t$ denote an exogenous instrument for a shock of interest. Adding $\zeta_t$ to our general nonlinear regression problem and iterating $y_t$ $h$-periods forward (for $h=0,\dots, H$), yields:
\begin{equation} \label{eq:LP1}
    y_{t+h} =  \psi_h \zeta_t + \bm x_t' \bm \gamma_h + \bm \epsilon'_{t+h}  \tilde{\bm \gamma}_h + f_h(\zeta_t, \bm x_t, \bm \epsilon_{t+h}) + \varepsilon_{t+h}. 
\end{equation}
Here, for $h \ge 1$ we let  $\bm \epsilon_{t+h} = (\varepsilon_{t}, \dots, \varepsilon_{t+h-1})'$ denote a $h$-dimensional vector of shocks for periods  $t,\dots, t+h-1$, and $ \tilde{\bm \gamma}_h$ a $h$-dimensional vector of associated coefficients. For $h=0$, no shocks are included. Note that $\psi_h, \bm \gamma_h, \tilde{\bm \gamma}_h$ and $f_h$ are horizon-specific. 

We obtain the nonlinear local projections in two steps \citep[see, e.g.,][]{kilian2017structural}. First, we obtain the NLP conditional on the full history of the data $\bm \Omega_{t}$ up to time $t$ (which includes the instrument, past shocks and covariates) by computing the difference between the expectation of Eq. (\ref{eq:LP1}) given $\zeta_t = \tau$  and the expectation given $\zeta_t = 0$:
\begin{equation*}
    \text{NLP}(h, \tau, \bm \Omega_{t}) = \mathbb{E}(y_{t+h}|\zeta_t=\tau, \bm \Omega_{t})  - \mathbb{E}(y_{t+h}|\zeta_t=0, \bm \Omega_{t}).
\end{equation*}
Second, this NLP depends on the observed data $\bm \Omega_{t}$. A more general way of representing the nonlinear response of the economy to a financial shock can be obtained by considering the unconditional NLP obtained as:
\begin{equation}
    \text{NLP}(h, \tau) = \int \text{NLP}(h, \tau, \bm \Omega^r_{t})  d\bm \Omega^r_{t}, \label{eq:NLP_cond}
\end{equation}
 where, as in \cite{kilian2017structural}, $\bm \Omega^r_{t}$ is a randomly selected path of observations. 
 
 This quantity can be computed for each MCMC draw  of $\psi_h, \bm \gamma_h, \tilde{\bm \gamma}_h $ and $\widehat{f}_h$ (the trained BNN approximation to $f_h$), yielding a posterior distribution over impulse responses to financial shocks. Specifically, for each MCMC draw we compute NLP$(h, \tau, \bm \Omega_{t})$ for $R$ different realizations of $\bm \Omega_{t}^r$ and take the mean:
 \begin{equation}
     \text{NLP}(h, \tau) \approx \frac{1}{R} \sum_{r=1}^R  \text{NLP}(h, \tau, \bm \Omega^r_{t}).
 \end{equation}
 Setting $R = 400$ to a large value yields a precise approximation to the integral in Eq. (\ref{eq:NLP_cond}).

An additional complication to compute NLP($h, \tau$) is that $\bm \epsilon_{t+h}$ in Eq. (\ref{eq:LP1}) is latent.\footnote{In an extensive robustness check, \cite{clark2024investigating} show that of  $\bm \epsilon_{t+h}$ only has a small impact on direct forecasts.} Within a linear framework, \cite{lusompa2023local} proposes  estimating $\bm \epsilon_{t+h}$ using the estimated residuals and then treating them as fixed regressors.

Yet, in a nonlinear framework, ignoring uncertainty surrounding  $\bm \epsilon_{t+h}$ can generate bias. Hence, we follow an alternative, sequential approach which exploits the Bayesian nature of our method. More specifically, in a first step we estimate the regression in Eq. (\ref{eq:LP1}) for $h=0$, save the posterior distribution of the shock $\varepsilon_t$. Second, we estimate the model for $h=1$, replacing $\epsilon_{t+1}$ in each draw of the Gibbs sampler by a draw $\epsilon^{(j)}_{t+1}=\varepsilon^{(j)}_t$ from $p(\epsilon_{t+1}|\bullet)$ for each $t$. This yields a shock distribution of $p(\varepsilon_{t+1}|\bullet)$. Third, we estimate the model for $h=2$ replacing  $\bm \epsilon_{t+2}$ by $\bm \epsilon^{(j)}_{t+2} = (\varepsilon^{(j)}_t, \varepsilon^{(j)}_{t+1})' \sim p(\varepsilon^{(j)}_t, \varepsilon^{(j)}_{t+1}|\bullet)$ for each $t$. This procedure is repeated until we end up estimating the regression for horizon $H$. 

\section{The nonlinear effects of financial shocks}\label{seclp:results}
We focus on how financial shocks impact inflation, output and employment, and whether these effects are symmetric and proportional. To answer these questions, we employ a BNN featuring one hidden layer and $Q=K$ neurons. Building on the existing literature \citep{gilchrist2012credit,barnichon2022effects,mumtaz2022impulse}, we construct our dataset using variables from FRED-MD \citep{mccracken2016fred}, including inflation (CPIAUCSL), employment (CE16OV), industrial production growth (INDPRO) as well as the federal funds rate (FEDFUNDS) and stock market returns (S.P.500). In addition, we also include a proxy of the financial shock.  This proxy is obtained by including the excess bond premium \citep[EBP,][]{gilchrist2012credit} in a structural VAR that uses the same five variables as well as the EBP. We order the EBP measure first and obtain the structural economic shock related to this variable, which can be interpreted as a financial shock. The sample ranges from January 1960 to December 2020.

\begin{figure}[th!]
\caption{Nonlinear local projections to financial shocks. \label{figlp:shlwBNN_lp}}
\centering
\begin{minipage}{0.49\textwidth}
\centering
\normalsize (a) \textit{Sign asymmetries}
\end{minipage}
\begin{minipage}{0.49\textwidth}
\centering
\normalsize (b) \textit{Size asymmetries}
\end{minipage}
\begin{minipage}{1\textwidth}
\centering
\vspace*{10pt}
\small \textit{Inflation}
\vspace*{10pt}
\end{minipage}
\begin{minipage}{0.49\textwidth}
\centering
\includegraphics[scale=.55]{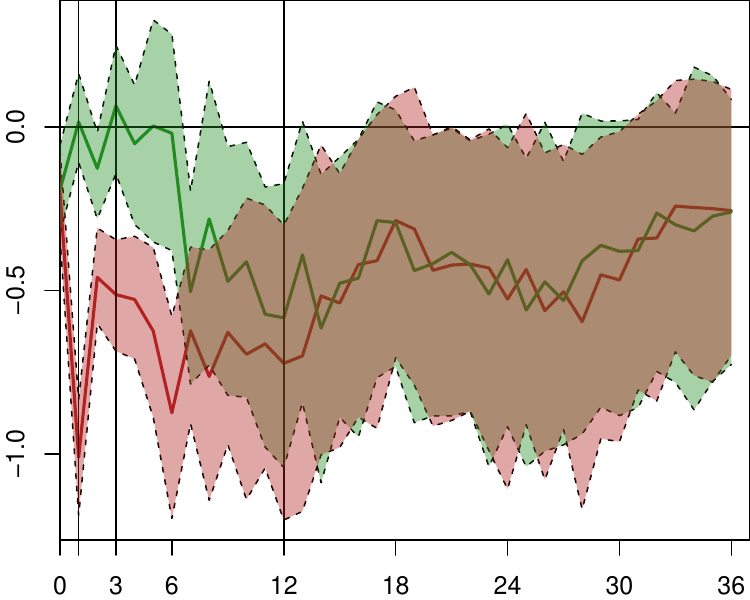}
\end{minipage}
\begin{minipage}{0.49\textwidth}
\centering
\includegraphics[scale=.55]{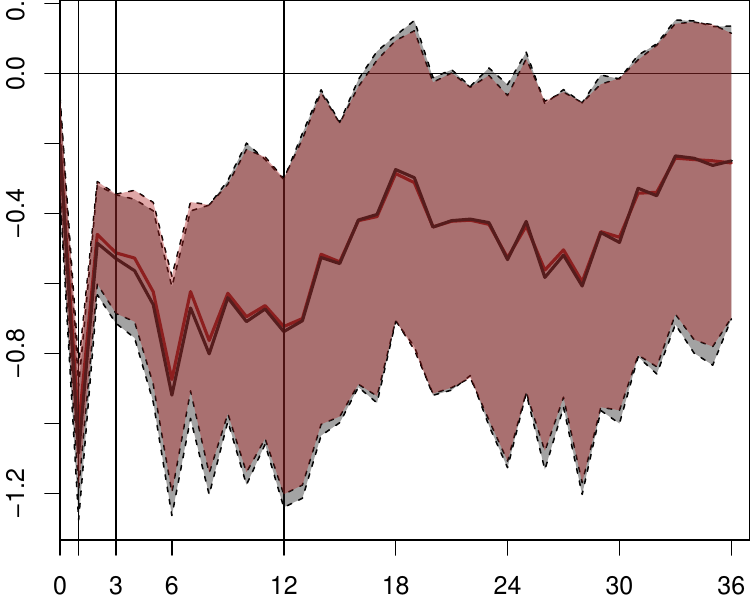}
\end{minipage}
\begin{minipage}{1\textwidth}
\centering
\vspace*{10pt}
\small \textit{Industrial production}
\vspace*{10pt}
\end{minipage}
\begin{minipage}{0.49\textwidth}
\centering
\includegraphics[scale=.55]{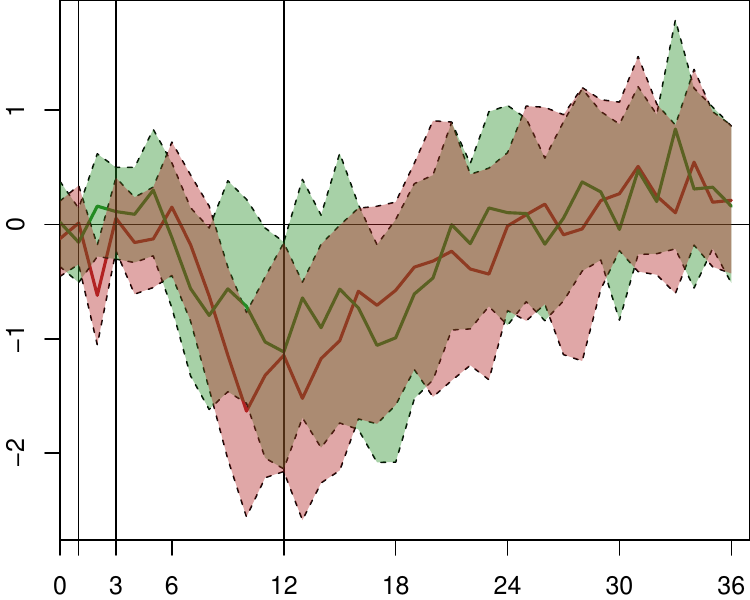}
\end{minipage}
\begin{minipage}{0.49\textwidth}
\centering
\includegraphics[scale=.55]{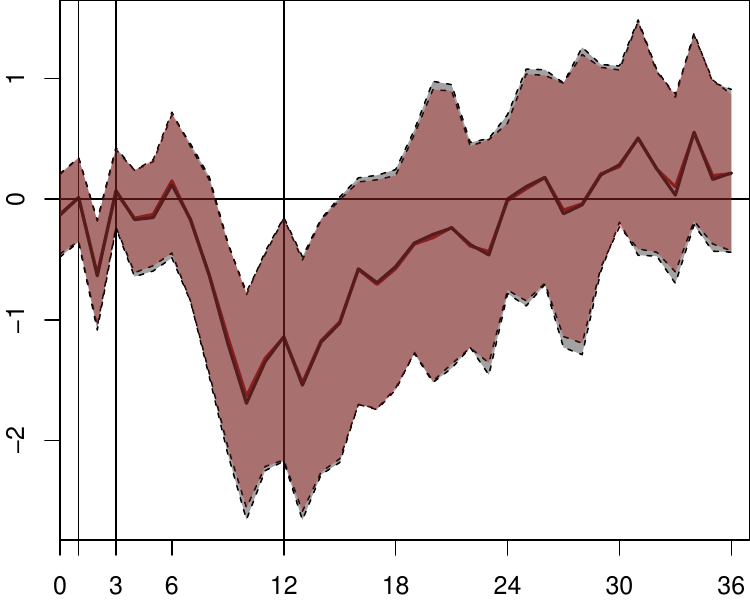}
\end{minipage}
\begin{minipage}{1\textwidth}
\centering
\vspace*{10pt}
\small \textit{Employment}
\vspace*{10pt}
\end{minipage}
\begin{minipage}{0.49\textwidth}
\centering
\includegraphics[scale=.55]{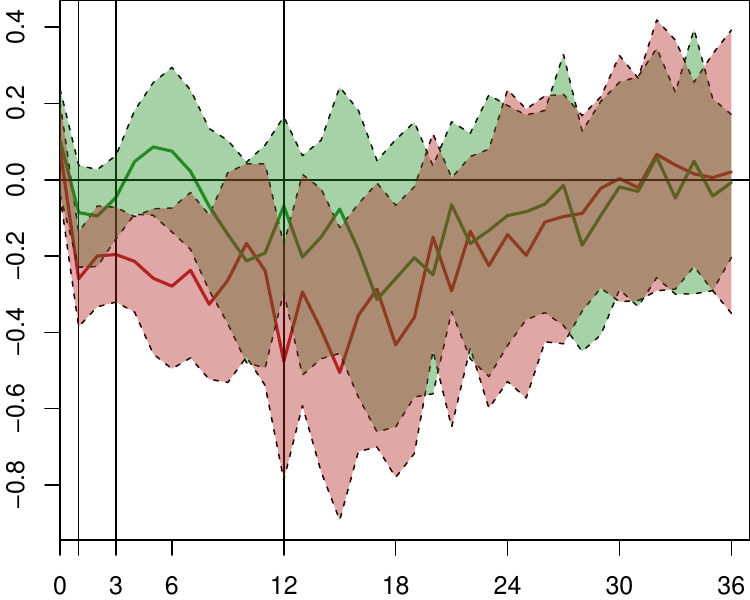}
\end{minipage}
\begin{minipage}{0.49\textwidth}
\centering
\includegraphics[scale=.55]{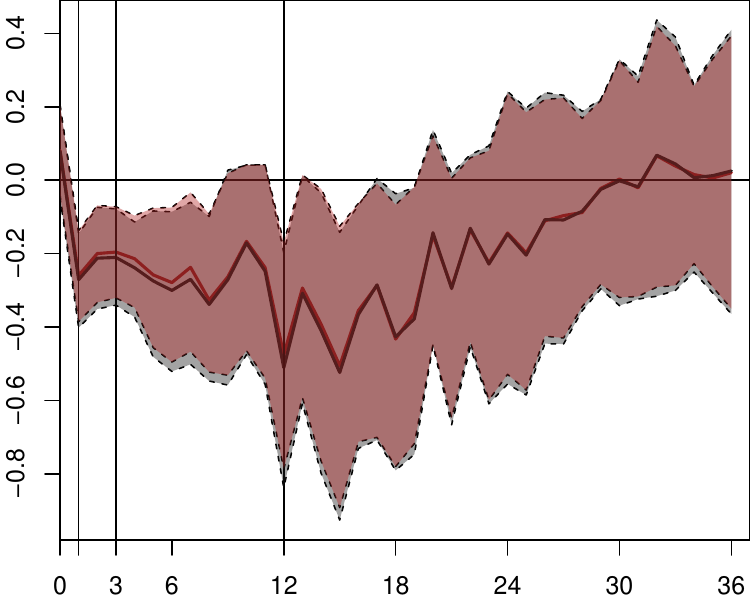}
\end{minipage}
\begin{minipage}{\textwidth}
\vspace{5pt}
\scriptsize \emph{Note:} This figure shows asymmetries in local projection responses to shocks to the excessive bond premium (EBP). The left-hand panel refers to asymmetries in the sign of the shock and the right-hand panel refers to asymmetries in the size of the shock. The solid lines indicate the posterior median, while the shaded areas refer to the $68\%$ posterior credible interval.
A positive one-unit shock is denoted in red, a negative one-unit shock is shown in green, and a positive three-unit shock is shown in gray. To ease comparability, we re-scale the responses associated with a negative shock and the responses associated with three-unit shock such that they represent a positive one-unit shock (multiplied by $-1$ and $1/3$, respectively). 
\end{minipage}
\end{figure}

Figure \ref{figlp:shlwBNN_lp} shows the 16$^{th}$ and 84$^{th}$ percentiles of the posterior distribution of the LPs (shaded areas) and the median (solid lines)  for different shock signs and shock sizes. Panel (a) shows the responses of inflation, IP and employment to a one unit contractionary financial shock (in red) and to a benign financial shock (in green). The benign shock is multiplied by $-1$ to simplify comparison. Panel (b) shows the responses to a contractionary financial shock to a one and three unit financial shock. The three unit shock is re-scaled by $1/3$ to ease comparison.

The figure reveals substantial asymmetries with respect to the sign of the shock. Consistent with the literature \citep[see, e.g., ][]{barnichon2016theory,barnichon2022effects,forni2024nonlinear}, we find that contractionary shocks have much stronger effects on the macro aggregates than benign shocks. Starting with inflation reactions, we find that contractionary financial shocks exert downward pressure on prices. This effect is strong and peaks after around one month, with a  peak median decline in inflation of around one percentage point. Notice that the reaction is also quite persistent and turns insignificant only after around 18 months. By contrast, a benign shock leads to a much weaker reaction of inflation. After around seven months (until around 1.5 years), there is some evidence that inflation picks up. But apart from this, the credible intervals mostly include zero. 

We now turn to the reaction of IP growth. Output seems to react sluggishly with respect to financial shocks. The declines in output growth are much stronger if the shock is adverse, reaching almost two percentage points after around eleven months. For a benign shock, the effects are much more muted and (almost) never significant.

Similar to the inflation reaction, we find that employment  growth strongly declines after a negative financial shock. This decline peters out after around two years, turning insignificant afterwards. The sharp drop in prices can be linked to the decline in employment and the associated downward pressure on wages. Again, we find no discernible reaction to a benign shock.

In terms of size asymmetries (panel (b) of Figure \ref{figlp:shlwBNN_lp}), we find that contractionary financial shocks of different sizes (one and three units) trigger proportional reactions of all three focus variables under consideration.

\section{Conclusion}\label{seclp:conclusion}
We exploit Bayesian neural networks to compute non-parametric impulse response functions, and applies the method to approximate the reaction of three ley US macroeconomic variables to financial shocks. We find that asymmetries arise mostly with respect to the sign of the shocks. Adverse financial shocks have a tendency to trigger much stronger reactions of inflation, industrial production and employment than benign shocks. When it comes to asymmetries with respect to size, we instead find no differences, with small and large shocks resulting in almost exactly proportional impulse responses.

\clearpage
\small{\setstretch{0.85}
\addcontentsline{toc}{section}{References}
\bibliographystyle{cit_econometrica.bst}
\bibliography{lit}}

@article{hauzenberger2024BNN,
  title={Bayesian neural networks for macroeconomic analysis},
  author={Hauzenberger, Niko and Huber, Florian and Klieber, Karin and Marcellino, Massimiliano},
  journal={Journal of Econometrics},
  volume={forthcoming},
  year={2024},
  publisher={Elsevier}
}

@book{kilian2017structural,
  title={Structural vector autoregressive analysis},
  author={Kilian, Lutz and L{\"u}tkepohl, Helmut},
  year={2017},
  publisher={Cambridge University Press}
}

@article{lusompa2023local,
  title={Local projections, autocorrelation, and efficiency},
  author={Lusompa, Amaze},
  journal={Quantitative Economics},
  volume={14},
  number={4},
  pages={1199--1220},
  year={2023},
  publisher={Wiley Online Library}
}

@article{clark2024investigating,
  title={Investigating Growth-at-Risk Using a Multicountry Non-parametric Quantile Factor Model},
  author={Clark, Todd E and Huber, Florian and Koop, Gary and Marcellino, Massimiliano and Pfarrhofer, Michael},
  journal={Journal of Business \& Economic Statistics},
  volume={forthcoming},
  year={2024},
  publisher={Taylor \& Francis}
}

@article{inoue2024local,
  title={Local projections in unstable environments},
  author={Inoue, Atsushi and Rossi, Barbara and Wang, Yiru},
  journal={Journal of Econometrics},
  pages={105726},
  year={2024},
  publisher={Elsevier}
}

@article{gonccalves2024state,
  title={State-dependent local projections},
  author={Gon{\c{c}}alves, S{\'\i}lvia and Herrera, Ana Mar{\'\i}a and Kilian, Lutz and Pesavento, Elena},
  journal={Journal of Econometrics},
  volume={105702},
  year={2024},
  publisher={Elsevier}
}

@article{jorda2005estimation,
  title={Estimation and inference of impulse responses by local projections},
  author={Jord{\`a}, {\`O}scar},
  journal={American Economic Review},
  volume={95},
  number={1},
  pages={161--182},
  year={2005},
  publisher={American Economic Association}
}

@inproceedings{carvalho2009handling,
  title={Handling sparsity via the horseshoe},
  author={Carvalho, Carlos M and Polson, Nicholas G and Scott, James G},
  booktitle={Artificial intelligence and statistics},
  pages={73--80},
  year={2009},
  organization={PMLR}
}

@article{bhadra2020horseshoe,
  title={Horseshoe regularisation for machine learning in complex and deep models},
  author={Bhadra, Anindya and Datta, Jyotishka and Li, Yunfan and Polson, Nicholas},
  journal={International Statistical Review},
  volume={88},
  number={2},
  pages={302--320},
  year={2020},
  publisher={Wiley Online Library}
}

@article{ghosh2019model,
  title={Model selection in Bayesian neural networks via horseshoe priors},
  author={Ghosh, Soumya and Yao, Jiayu and Doshi-Velez, Finale},
  journal={Journal of Machine Learning Research},
  volume={20},
  number={182},
  pages={1--46},
  year={2019}
}

@article{mumtaz2022impulse,
  title={Impulse response estimation via flexible local projections},
  author={Mumtaz, Haroon and Piffer, Michele},
  journal={arXiv:2204.13150},
  year={2022}
}

@article{gilchrist2012credit,
  title={Credit spreads and business cycle fluctuations},
  author={Gilchrist, Simon and Zakraj{\v{s}}ek, Egon},
  journal={American Economic Review},
  volume={102},
  number={4},
  pages={1692--1720},
  year={2012},
  publisher={American Economic Association}
}

@article{barnichon2022effects,
  title={Are the effects of financial market disruptions big or small?},
  author={Barnichon, Regis and Matthes, Christian and Ziegenbein, Alexander},
  journal={Review of Economics and Statistics},
  volume={104},
  number={3},
  pages={557--570},
  year={2022},
  publisher={MIT Press One Rogers Street, Cambridge, MA 02142-1209, USA journals-info~…}
}

@article{kastner2014ancillarity,
  title={Ancillarity-sufficiency interweaving strategy (ASIS) for boosting MCMC estimation of stochastic volatility models},
  author={Kastner, Gregor and Fr{\"u}hwirth-Schnatter, Sylvia},
  journal={Computational Statistics \& Data Analysis},
  volume={76},
  pages={408--423},
  year={2014},
  publisher={Elsevier}
}

@article{mccracken2016fred,
  title={FRED-MD: A monthly database for macroeconomic research},
  author={McCracken, Michael W and Ng, Serena},
  journal={Journal of Business \& Economic Statistics},
  volume={34},
  number={4},
  pages={574--589},
  year={2016},
  publisher={Taylor \& Francis}
}

@article{makalic2015simple,
  title={A simple sampler for the horseshoe estimator},
  author={Makalic, Enes and Schmidt, Daniel F},
  journal={IEEE Signal Processing Letters},
  volume={23},
  number={1},
  pages={179--182},
  year={2015},
  publisher={IEEE}
}

@article{neal2011hmc,
  title={MCMC using Hamiltonian dynamics},
  author={Neal, Radford M},
  journal={Handbook of Markov Chain Monte Carlo},
  volume={2},
  number={11},
  pages={2},
  year={2011},
  publisher={Chapman and Hall/CRC}
}

@article{barnichon2016theory,
  title={Theory ahead of measurement? assessing the nonlinear effects of financial market disruptions},
  author={Barnichon, Regis and Matthes, Christian and Ziegenbein, Alexander},
  year={2016},
  journal={FRB Richmond Working Paper}
}

@article{forni2024nonlinear,
  title={Nonlinear transmission of financial shocks: Some new evidence},
  author={Forni, Mario and Gambetti, Luca and Maffei-Faccioli, Nicol{\`o} and Sala, Luca},
  journal={Journal of Money, Credit and Banking},
  volume={56},
  number={1},
  pages={5--33},
  year={2024},
  publisher={Wiley Online Library}
}

@article{balke2000credit,
  title={Credit and economic activity: credit regimes and nonlinear propagation of shocks},
  author={Balke, Nathan S},
  journal={Review of Economics and Statistics},
  volume={82},
  number={2},
  pages={344--349},
  year={2000},
  publisher={MIT Press 238 Main St., Suite 500, Cambridge, MA 02142-1046, USA journals~…}
}

@article{brunnermeier2014macroeconomic,
  title={A macroeconomic model with a financial sector},
  author={Brunnermeier, Markus K and Sannikov, Yuliy},
  journal={American Economic Review},
  volume={104},
  number={2},
  pages={379--421},
  year={2014},
  publisher={American Economic Association 2014 Broadway, Suite 305, Nashville, TN 37203}
}



\end{document}